\documentstyle[prl,twocolumn,aps,epsfig]{revtex}
 \draft

\begin{document}

\title{ Collective oscillations of a 1D trapped Bose gas}
 
\author{Chiara Menotti$^{1,2}$ and Sandro Stringari$^1$}
\address{$^1$ Dipartimento di Fisica, Universit\`{a} di Trento,}
\address{and Istituto Nazionale per la Fisica della Materia,}
\address{I-38050 Povo, Italy}
\address{$^2$ Universit\`{a} Cattolica del Sacro Cuore,}
\address{I-25121 Brescia, Italy}

  \date{\today}

  \maketitle
\begin{abstract}
\noindent
Starting from the hydrodynamic equations of superfluids,
we calculate the  frequencies of the collective oscillations of a harmonically 
trapped Bose gas for various 1D configurations. These include the
mean field regime described by  Gross-Pitaevskii theory and the beyond mean 
field regime at small  densities described by Lieb-Liniger theory.  
The relevant combinations of the physical parameters governing 
the transition between the different regimes are  discussed.

\end{abstract}

 
 \narrowtext

Recent experiments on trapped Bose gases at low temperature have pointed out
the occurrence of characteristic 1D features. These include deviations of the
aspect ratio and of the release energy \cite{mit01,salomon} from the 
3D behaviour as well as the appearence of thermal fluctuations of the 
phase, peculiar of 1D configurations \cite{hannover}.
Interest in 1D interacting Bose gases arises from the occurrence of  
quantum features which are not encountered in 2D and 3D. For example  
in 1D the  fluctuations of the phase  of the order parameter
rule out the occurrence of long range order even at zero 
temperature \cite{PS}. 
Such systems cannot be in general described using traditional
mean field theories and require the development of a 
more advanced many-body approach. 
In the case of  1D Bose gases interacting with repulsive zero-range 
forces, this has  been implemented  by Lieb and Liniger \cite{lieb} who 
studied both the equation of state and the spectrum of elementary 
excitations of a uniform gas. In the presence of harmonic trapping,
1D Bose gases exhibit new interesting features. The corresponding 
equilibrium properties have been already discussed  in a recent series 
of theoretical papers (see  \cite{gora,olshanii,gir} and references therein). 
In the present work we investigate the consequences of  harmonic trapping  
on the collective oscillations of an interacting 1D Bose gas at zero 
temperature.
We will consider various configurations, ranging from the mean field 
regime \cite{rmp} where the healing length is larger than the average 
interparticle distance, to  the Tonks-Girardeau limit \cite{tonks} where 
the gas acquires Fermi like properties.   
We will show that  the   frequency of the lowest compression mode provides
 a useful indicator of the different regimes.

We start our discussion from the hydrodynamic equations of superfluids in 1D
\begin{eqnarray}
\label{HD1}
&&\frac{\partial}{\partial t}\delta n_1 + 
\frac{\partial}{\partial z} \left(n_1 v \right) =0 \;, \\
&&\frac{\partial}{\partial t} v + 
\nabla_z \left(\mu_{\ell e}(n_1) +V_{ext}+\frac{1}{2}mv^2 \right) =0 \;,
\label{HD2}
\end{eqnarray}
which decribe  the dynamic behaviour of such systems at zero temperature.
In these equations $n_1(z,t)$ is the 1D  density of the gas, $v(z,t)$ is the 
velocity field, while $V_{ext}(z)$ is the external trapping potential 
which in the following will be assumed to be harmonic: 
$V_{ext}(z) = m\omega_z^2z^2/2$. 
The hydrodynamic equations of superfluids have been already
successfully employed to predict the collective frequencies
of 3D trapped Bose-Einstein condensates \cite{sandro96}.
A crucial ingredient of these equations is the local equilibrium
($\ell e$) chemical potential $\mu_{\ell e}$, which should be 
evaluated for a uniform 1D gas 
($V_{ext}=0$) at the density $n_1$. 
The applicability of the above equations 
requires the validity of the local density approximation along the 
$z$-th direction. This is expected to be accurate for 
sufficiently large systems.
Furthermore, Eqs.(\ref{HD1},\ref{HD2}) 
should be limited to the study of macroscopic 
phenomena where variations in space take place over distances
larger than the average distance between particles.
From Eq.(\ref{HD2}) one can easily calculate 
the ground state profile through the equation
\begin{equation}
\mu_{\ell e}(n_1(z)) + V_{ext}(z) = \mu \; .
\label{n0}
\end{equation}
The collective oscillations are instead 
determined by writing 
the density in the form $n_1(z,t) = n_1(z) + e^{-i\omega t}\delta n_1(z)$, 
with the function $\delta n_1(z)$ obeying the
linearized equation 
\begin{equation}
\omega^2 \delta n_1(z) = 
\frac{1}{m}\nabla_z \left[ n_1(z) \nabla_z \left( 
\frac{\partial \mu_{\ell e}}{\partial n_1} \delta n_1(z) \right)
\right] \;,
\label{deltan}
\end{equation}
which immediately follows from Eqs.(\ref{HD1},\ref{HD2}).
 In the case of  uniform systems ($V_{ext}=0$) one has 
plane wave solutions $\delta n_1 = \exp(iqz)$ with $\omega^2 = c^2q^2$,
where $q$ is the wave vector of the excitation and 
$c^2 =n_1 (\partial \mu_{\ell e}  /\partial n_1) /m$ 
is the square of the sound velocity. It is worth reminding that 
the applicability of the hydrodynamic equations is not limited 
to the mean field scenario. Actually in \cite{lieb} it has been proven
that in 1D  Bose gases the velocity of sound,
derived from the macroscopic compressibility, coincides with the one derived 
from the microscopic calculation of the phonon excitation spectrum
also in regimes far from mean field.

In the first part of the work we evaluate $\mu_{\ell e}(n_1)$ and the 
corresponding solutions for the collective oscillations in the framework 
of the mean field Gross-Pitaevskii theory.
Even  in this regime one can 
explore a  rich variety of situations ranging from the Thomas-Fermi regime 
in the radial direction to the one of tight confinement where the motion 
in the radial direction is frozen.
In the second part we extend the analysis to 
regimes beyond mean field, including the limit of the Tonks-Girardeau gas.

Let us  consider  a uniform system of length $L$ in the $z$-th direction 
and confined by a harmonic potential 
$V(r_{\perp})=m\omega_{\perp}^2 r_{\perp}^2/2$ in the radial direction. 
By writing the order parameter in the form 
$\Psi=\sqrt{n_1}f(\rho_{\perp})/a_{\perp}$, where $n_1 =N/L$ is the 1D 
density, $a_{\perp}=\sqrt{\hbar / m\omega_{\perp}}$ is the 
oscillator length in the radial direction and 
$\rho_{\perp}=r_{\perp}/a_{\perp}$ is the dimensionless radial 
coordinate, the 3D Gross-Pitaevskii equation yields the dimensionless 
equation 
\begin{equation}
\left(-\frac{1}{2} \frac{\partial^2}{\partial \rho_{\perp}^2}
-\frac{1}{2\rho_{\perp}}  \frac{\partial}{\partial \rho_{\perp}}
+ \frac{1}{2} \rho_{\perp}^2 + 4 \pi a n_{1} f^2 \right) f = 
\frac{ \mu_{\ell e}}{\hbar \omega_{\perp}} f
\label{f}
\end{equation}
for the function $f$ obeying the normalization 
condition $2\pi \int |f(\rho_{\perp})|^2 \rho_{\perp} d \rho_{\perp} =1$. 
In Eq.(\ref{f}), 
$\mu_{\ell e} /\hbar \omega_{\perp}$ is the chemical 
potential in units of the radial quantum oscillator energy.
Eq. (\ref{f}) shows that the relevant dimensionless parameter of 
the problem  is $an_1$. 
It is worth considering two important limits. 
If $a n_1 \gg 1$, one  enters the  radial Thomas-Fermi regime 
(hereafter called 3D cigar), where many configurations
of the harmonic oscillator Hamiltonian are excited in the radial 
direction and the equation of state takes the analytic form 
\begin{equation}
{\mu_{\ell e} \over \hbar \omega_{\perp}} = 2 (a n_1)^{1/2}.
\label{mu3D}
\end{equation}
Notice that in this limit  the chemical potential is not linear 
in the density. This implies, in particular, that the sound velocity 
is related to the chemical potential by the law 
$c^2 =\mu_{\ell e}/2m$ \cite{zaremba} rather then by the
Bogoliubov relation $c^2 =\mu_{\ell e}/m$.
A second important case is the perturbative regime where $a n_1 \ll 1$ 
(hereafter called 1D mean field).
In this case the solution of (\ref{f}) approaches the Gaussian ground 
state of the radial harmonic oscillator  and one finds  the linear 
law
\begin{equation}
{\mu_{\ell e} \over \hbar \omega_{\perp}} = 1 + 2 a n_1
\label{mu1D}
\end{equation}
for the chemical potential.

If the gas is harmonically trapped also along the $z$-th direction, 
the density $n_1$ exhibits a $z$-th dependence which is worth calculating
as a function of the 
relevant parameters of the problem: the scattering length $a$, the 
number $N$ of atoms and the trapping frequencies 
$\omega_{\perp}$ and $\omega_z$. 
To this purpose one has to solve Eq.(\ref{n0}) by imposing the normalization
condition $\int n_1(z) dz=N$. 
A useful quantity is the Thomas-Fermi radius $Z$ defined by the value of $z$
at which the equilibrium  density $n_1(z)$ vanishes.
According to Eq.(\ref{n0}), one has 
$\mu-\mu_{\ell e}(an_1=0) = (1/2)m\omega_z^2Z^2$.
In terms of $Z$,  Eq.(\ref{n0}) can be rewritten as 
${\tilde \mu}_{\ell e}(an_1(z)) = 
(m\omega_z^2 Z^2/2\hbar\omega_{\perp})(1-z^2/Z^2)$, 
where we have defined the dimensionless quantity
$ {\tilde \mu}_{\ell e}(an_1(z))=
\left[\mu_{\ell e}(an_1(z))- \mu_{\ell e}(an_1=0)\right]/\hbar \omega_{\perp}$.
This function is fixed by the solution of the Gross-Pitaevskii equation
(\ref{f}).
Its  inverse ${\tilde \mu}_{\ell e}^{-1}$ gives
the value of $a n_1$ as a a function of $z$, so that the normalization 
condition obeyed by the density can be written as
\begin{equation}
\frac{Za_{\perp}}{a_z^2} 
\int_{-1}^{1} {\tilde \mu}_{\ell e}^{-1} 
\left[ \frac{1}{2} 
\left( \frac{Za_{\perp}}{a_z^2} \right)^{2} (1-t^2) \right] dt=
\frac{N a a_{\perp}}{a_z^2} ,
\label{s5}
\end{equation}
with $t=z/Z$. 
Eq. (\ref{s5}) explicitly points out  the relevance of 
the dimensionless combination $Naa_{\perp}/a^2_z$ where 
$a_z =\sqrt{\hbar / m\omega_z}$ is the oscillator length in the
axial directions.
From Eq.(\ref{s5}) one can calculate, for a given choice of 
the parameters, the radius $Z$ and hence the $1D$ density profile. 
In the 3D cigar limit $Naa_{\perp}/a^2_z\gg 1$, one has
$Z =  (a^2_z/a_{\perp})(15 Naa_{\perp}/a^2_z)^{1/5}$ and

\begin{eqnarray}
n_1(z) = \frac{1}{16 a}
\left( \frac{15 Naa_{\perp}}{a^2_z}\right)^{4/5}
\left(1-\frac{z^2}{Z^2}\right)^2.
\end{eqnarray}
In the 1D mean field limit $Naa_{\perp}/a^2_z\ll 1$,
one instead finds \cite{olshanii}
$Z= (a^2_z/a_{\perp})(3Naa_{\perp}/a^2_z)^{1/3}$ and

\begin{eqnarray}
n_1(z) = \frac{1}{4a}
\left(\frac{3Naa_{\perp}}{a^2_z}\right)^{2/3} 
\left(1-\frac{z^2}{Z^2}\right).
\label{n1MF}
\end{eqnarray}
The density profiles are 
different in the two regimes, reflecting the different behaviour of 
the equation of state.
The conditions of applicability of the local density approximation 
employed above are determined by requiring that $Z \gg a_z$. In the 
1D mean field regime  this implies the non trivial condition
$(a_z/a_{\perp}) (N a a_{\perp}/a_z^2)^{1/3} \gg 1$.

Let us now discuss the behaviour of the collective oscillations. 
The hydrodynamic equation (\ref{deltan}) has simple analytic solutions if 
the density derivative of the chemical potential is a power law function:
$\partial \mu_{\ell e} /\partial n_1 \propto n_1^{\gamma-1}$. 
The Thomas-Fermi  and the 1D mean field regimes belong to this class 
of solutions with $\gamma =1/2$ and $\gamma =1$ respectively. 
Also the Tonks-Girardeau limit (see Eq.(\ref{muT}) below) belongs to 
the same class with 
$\gamma = 2$. Hence in these three relevant limits the dipersion relation 
for the collective frequencies can be obtained analytically. By looking 
for solutions of the form 
$n_1^{\gamma -1}\delta n_1(z) = z^k + a z^{k-2} \dots$, where 
$k \le 1$ and only positive powers of $z$ are included in the
polynomial, one finds the result
\begin{equation}
\omega^2 = \omega^2_z {k \over 2} [2 +\gamma (k-1)].
\label{omegagamma}
\end{equation}
The case $k=1$ corresponds to the center of mass motion whose frequency 
is given by $\omega=\omega_z$ independent of the value of $\gamma$. 
The most interesting $k=2$ case (lowest compressional mode) is instead 
sensitive to the regime considered. 
One finds $\omega^2 = (5/2)\omega_z^2$, $\omega^2 = 3\omega_z^2$ and 
$\omega^2 = 4\omega_z^2$ for the 3D cigar, 1D mean field and Tonks-Girardeau 
regimes respectively. The result $\omega^2 = (5/2)\omega_z^2$
was first derived in \cite{sandro96} by solving the hydrodynamic 
equations for a trapped 3D system in the limit of a highly elongated
trap ($\omega_z \ll \omega_{\perp}$). 
This prediction has been confirmed experimentally with high precision 
\cite{mit96}. 
The result $\omega^2 = 3 \omega^2_z$ was derived in 
\cite{csordas,sandro98,ho}, 
while the result  $\omega^2 = 4\omega^2_z$ for the Tonks-Girardeau gas 
is simply understood by recalling that, in this limit, there is an exact 
mapping with the 1D ideal Fermi gas \cite{tonks} where the excitation 
spectrum, in the presence of harmonic confinement, is  $\omega = k \omega_z$.  
The same result has been recently derived in
\cite{Minguzzi} using the 
mean field equations of Kolomeisky et al. \cite{kol}.

\begin{center}
\begin{figure}
\includegraphics[width=0.95\linewidth]{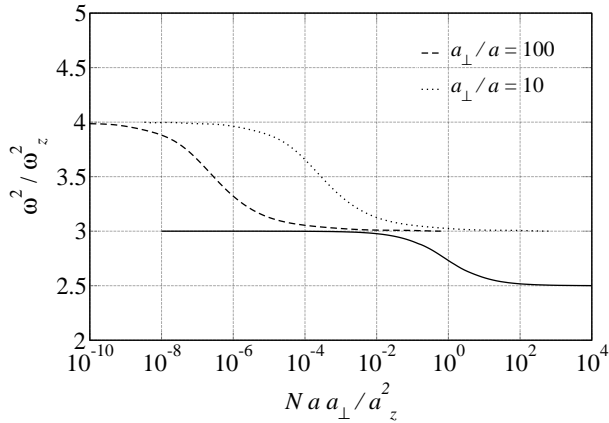}
\caption{Transition between the 1D mean field and the 3D cigar regimes: 
$\omega^2/\omega_z^2$ as a function of the parameter 
$Naa_{\perp}/a_z^2$ (full line). Transition to the Tonks-Girardeau regime for
$a_{\perp}/a=100$ (dashed line)
 and $10$ (dotted line).}
\label{fig1}
\end{figure}
\end{center}

In order to evaluate the collective frequencies in the intermediate 
regimes where 
the hydrodynamic equations are not  analytically  soluble, 
we have  developed a sum rule approach \cite{sandro96}, based on the 
evaluation of the ratio $\hbar^2 \omega^2 = m_1/m_{-1}$ between the 
energy weighted and inverse energy weighted sum rules. In the following 
we will limit the discussion to the lowest compression mode which is 
naturally excited by the operator $\sum_{i=1}^N z_i^2$. 
The energy weighted moment is given by 
$m_1= (1/2)\langle [\sum_{i=1}^N z_i^2,[H,\sum_{i=1}^N z_i^2]]\rangle=
 (2N\hbar^2/m)\langle z^2\rangle$ 
where $\langle z^2\rangle = \int n_1(z)z^2dz/N$ is the average square 
radius fixed by the ground state solution $n_1(z)$. 
The inverse energy weighted moment $m_{-1}$ is related to the static 
polarizability $\alpha$ by $m_{-1}= (1/2) \alpha$.
This  is evaluated by adding the pertubation $-\epsilon z^2$ 
to the Hamiltonian and calculating the corresponding changes 
$\delta \langle z^2\rangle $ of the expectation value of the
square radius:
$\alpha = N\delta\langle z^2\rangle /\epsilon$. 
Adding the perturbation  $-\epsilon z^2$ is equivalent 
to changing the frequency $\omega_z$ of the harmonic confinement, so 
that the result for the collective frequency takes the  compact
form \cite{note}
\begin{equation}
\omega^2 = -2 {\langle z^2\rangle \over 
d\langle z^2\rangle /d\omega^2_z}.
\label{omega1-1}
\end{equation}
We have calculated the ratio $\omega^2/\omega_z^2$ as a function of the 
dimensionless parameter $Naa_{\perp}/a^2_z$, thereby exploring the
transition between the 3D cigar and the 
1D mean field  regimes.
The results are reported in Fig.~\ref{fig1}. For the experimental
conditions of \cite{mit01,salomon} where 
$Naa_{\perp}/a^2_z=0.24$ and $0.08$, we 
predict $\omega^2/\omega^2_z=2.85$ and $2.91$ respectively, 
confirming that those experiments are actually touching 
the transition between the two mean field regimes. 

It is worth noticing that result (\ref{omega1-1}) for the collective 
frequency, being based on general sum rule arguments, applies also
to  regimes beyond mean field where the density profile 
$n_1(z)$ and hence the value of $\langle z^2\rangle$ cannot be evaluated 
starting from the Gross-Pitaevskii equation of state.
Deviations from the mean field regime become important when the 
healing length $\xi =(8\pi an)^{-1/2}$ is comparable to the average 
distance $d$ between particles. 
In  the presence of tight radial confinement one can use the 
relationship  $n=n_1/\pi a_{\perp}^2$ between the 3D density evaluated 
at $r_{\perp}=0$ and the 1D density  
$n_1=\int n(r_{\perp}) d{\vec r}_{\perp}$. 
When $a$ becomes smaller than $d$, one can write  $d=1/n_1$.
One then obtains the result
$\xi / d =  \sqrt{a^2_{\perp}n_1/8a }$
which becomes smaller and smaller 
as the 1D density decreases, thereby suggesting the occurrence
of important deviations from
the mean field behaviour for very dilute 1D samples. This
should be contrasted with the   3D case 
where the mean field condition  ($\xi > d$) is better
and better satisfied as the density decreases. 
The combination $a_{\perp}^2n_1/a$ can then be used as an indicator 
of the applicability of the mean field approach. When its value 
becomes  of the order of $1$ or smaller, one enters a new regime 
characterized by important quantum correlations.
The corresponding  many-body problem was investigated by
Lieb and Liniger \cite{lieb} who considered 1D repulsive zero-range 
potentials of the form $g_{1D}\delta(z)$. 
The interaction parameter $g_{1D}$  can be also written as 
$g_{1D}=\hbar^2/(ma_{1D}) $  where
 $a_{1D}$   is the one-dimensional scattering length \cite{olshanii2}. 
 By averaging the 3D interaction $4\pi^2\hbar^2a \delta({\bf r})/m$ 
over the radial density profile, one obtains the simple  identification 
$a_{1D}=a^2_{\perp}/a$ \cite{noteolshanii}. 
The comparison with the expression for $\xi/d$ shows that the 
deviations from the mean field increase
by decreasing  $a_{1D}n_1$.

In the Lieb-Liniger scenario the 
energy per particle $\epsilon(n_1)$,  
when  expressed in  units of the energy $\hbar^2 /2ma^2_{1D}$,
turns out to be a  universal function of the dimensionless parameter 
$ a_{1D}n_1$. 
Important limits  are the high density limit $a_{1D}n_1\gg 1$, where  
one finds $\epsilon(n_1) = \hbar^2 n_1/ma_{1D}$ and hence 
the mean field result (\ref{mu1D}) for the chemical potential 
$\mu_{\ell e}= \partial (n_1 \epsilon(n_1))/\partial n_1$  
(a part from the constant term arising from the radial external
force).  
Note that under the assumption $a_{\perp} \ll a$, the two conditions 
$an_1 \ll 1$ and $a_{1D}n_1 \gg 1$ required to realise the 1D mean 
field regime can be simultaneously satisfied.
The other important limit is the low density 
Tonks-Girardeau limit $a_{1D}n_1\ll 1$,
where  the chemical potential takes the value

\begin{eqnarray}
\mu_{\ell e} = \pi^2 \hbar^2 n_1^2 /2m. 
\label{muT}
\end{eqnarray}
Here the chemical potential  no longer depends
on the interaction coupling constant and reveals  a typical 
Fermi-like behaviour \cite{tonks}.

In the presence of axial harmonic
trapping   the ground state density profile has been evaluated 
in \cite{olshanii} using the local density approximation
(\ref{n0}).   In this case the normalization condition  
$\int n_1(z) dz=N$  takes the form

\begin{equation}
\frac{Za_{1D}}{a_z^2} 
\int_{-1}^{1} {\tilde \mu}_{\ell e}^{-1} 
\left[ \left( \frac{Za_{1D}}{a_z^2} \right)^{2} (1-t^2) \right] dt=
\frac{N a_{1D}^2}{a_z^2}.
\label{T5}
\end{equation}
Similarly to Eq.(\ref{s5}), 
we have introduced the  radius $Z$ at which the density vanishes and the
inverse of the function ${\tilde \mu}_{\ell e}(n_{1D}a_{1D})$,
where  ${\tilde \mu}_{\ell e}$ is now the chemical potential 
expressed in units of  $\hbar^2 /2ma^2_{1D}$.
Eq.(\ref{T5}) shows that the relevant combination of parameters 
is given by $N a_{1D}^2/a_z^2$. 
This differs by the factor $(a_{\perp}/a)^3$ from the combination 
$N a a_{\perp}^2/a_z^2$
characterizing the transition between the  
mean field regimes  discussed in the first part of this work.

\begin{center}
\begin{figure}
\includegraphics[width=0.95\linewidth]{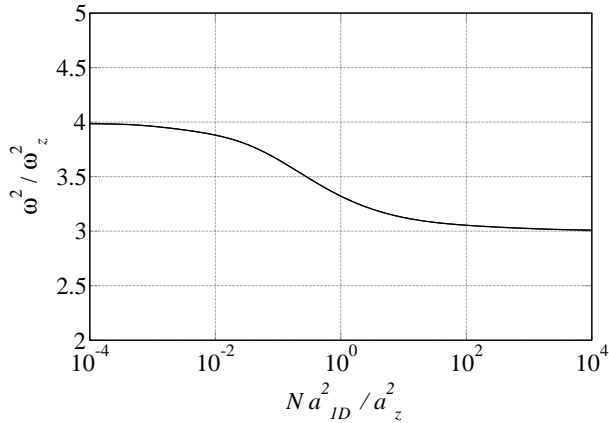} 
\caption{ Transition between the Tonks-Girardeau and the 1D mean field  
regimes: $\omega^2/\omega_z^2$ as a function of the parameter 
$Na_{1D}^2/a_z^2$.}
\label{fig2}
\end{figure}
\end{center}
Analytic solutions are obtained in the two limits $N a_{1D}^2/a_z^2 \gg 1$ 
and $N a_{1D}^2/a_z^2 \ll 1$. In the first case, one recovers the mean field
result (\ref{n1MF}). In the second one, we find the  profile \cite{gir,kol}

\begin{eqnarray}
n_1(z) = \frac{\sqrt{2N}}{\pi a_z}
\left(1-\frac{z^2}{Z^2}\right)^{1/2},
\end{eqnarray}
with $Z=\sqrt{2N}a_z$. 
In this case the  applicability of the local density approximation simply
requires $N \gg 1$.

By determining numerically the density profiles 
in the intermediate regimes we have calculated the frequency 
of the lowest compressional mode using the sum rule formula
(\ref{omega1-1}). 
The results are reported in Fig.~\ref{fig2} as a function of  
$Na_{1D}^2/a_z^2$. 
Fig.~\ref{fig1} shows the evolution of the collective frequency 
as a function of the parameter $Naa_{\perp}/a_z^2$ for two different 
choices of the ratio $a_{\perp}/a$. 
The corresponding curves  reveal the transition between 3D cigar,
1D mean field and Tonks-Giradeau regimes. 
It is however worth pointing out that  if the condition
$a_{\perp} \gg a$ is not well satisfied, the 1D mean field regime 
is not clearly identifiable 
 (see for example the dotted line in Fig.~\ref{fig1}).
In this case the determination of the collective
frequencies   requires 
the simultaneous  inclusion of 3D and beyond mean field effects.

This research is supported 
by the Ministero della Ricerca Scientifica e Tecnologica (MURST).

\noindent
%FIGURE CAPTION:

\noindent


\begin{references}
\bibitem{noteolshanii} The averaging procedure and the consequent 
identification of the 1D coup
ling constant is justified only if 
$a_{\perp} \gg a$ \cite{olshanii2}.
\end{references}
\end{document}